\documentclass[twocolumn,prd,nofootinbib]{revtex4}
\usepackage{amsmath}
\usepackage{amssymb}
\usepackage{graphicx}

\begin{document}

\title{E\"{o}tv\"{o}s branes}
\author{L\'{a}szl\'{o} \'{A}rp\'{a}d Gergely}
\affiliation{Department of Theoretical Physics, University of Szeged, Tisza Lajos krt
84-86, Szeged 6720, Hungary\\
Department of Experimental Physics, University of Szeged, D\'{o}m T\'{e}r 9,
Szeged 6720, Hungary}
\email{gergely@physx.u-szeged.hu}
\date{\today }

\begin{abstract}
The high value of brane tension has a crucial role in recovering Einstein's
general relativity at low energies. In the framework of a recently developed
formalism with variable brane tension one can pose the question, whether it
was always that high? In analogy with fluid membranes, in this paper we
allow for temperature dependent brane tension, according to the
corresponding law established by E\"{o}tv\"{o}s. For cosmological branes
this assumption leads to several immediate consequences: (a) The brane
Universe was created at a finite temperature $T_{c}$ and scale factor
\thinspace $a_{\min }$. (b) Both the brane tension and the 4-dimensional
gravitational coupling 'constant' increase with the scale factor from zero
to asymptotic values. (c) The 4-dimensional cosmological 'constant' evolves
with $a$, starting with a huge negative value, passing through zero, finally
reaching a small positive value. Such a scale--factor dependent cosmological
constant is able to generate a surplus of attraction at small $a$ (as dark
matter does) and a late-time repulsion at large $a$ (dark energy). In the
particular toy model discussed here the evolution of the brane tension is
compensated by energy interchange between the brane and the fifth dimension,
such that the continuity equation holds for the cosmological fluid. The
resulting cosmology closely mimics the standard model at late times, a
decelerated phase being followed by an accelerated expansion. The energy
absorption of the brane drives the 5D space-time towards maximal symmetry,
becoming Anti de Sitter.
\end{abstract}

\startpage{1}
\endpage{}
\maketitle

\section{Introduction}

Physics aims for a unified description of nature, tracing back all physical
laws to four fundamental interactions. While three of them are quantized,
and to certain extent further unified, gravity is still best described
classically. In contrast with the rest of the interactions about evolving
fields on a flat background, gravity is perceived as the dynamics of
geometry. String theory attempts to unify all interactions on different
grounds, its basic objects being open or closed strings and
higher-dimensional objects, called branes. The co-dimension one brane-world
theory (generalizing \ the early Randall-Sundrum model \cite{RS2}) carries
the original geometric spirit of general relativity, incorporating arbitrary
curvature and matter (for a review see \cite{MaartensLivRev}). The extra
dimension is both non-compact and curved (the remaining dimensions required
by string and M-theory can still be thought as compactified). Gravity acts
in 5-dimensions according to Einstein's equation with a negative
cosmological constant $\widetilde{\kappa }^{2}\widetilde{\Lambda }$, while
standard model fields are confined to the brane, a time-evolving
3-dimensional space-like hypersurface.

The projection of the 5-dimensional (5D) Einstein equation onto the brane
gives an effective Einstein equation, which in the most generic case reads 
\cite{Decomp}%
\begin{equation}
G_{ab}=-\Lambda g_{ab}+\kappa ^{2}T_{ab}+\widetilde{\kappa }^{4}S_{ab}-%
\overline{\mathcal{E}}_{ab}+\overline{L}_{ab}^{TF}+\overline{\mathcal{P}}%
_{ab}\ ,  \label{modEgen}
\end{equation}%
with $T_{ab}$, the brane energy-momentum tensor; $S_{ab}$, a quadratic
expression in $T_{ab}$; and $\overline{\mathcal{E}}_{ab}$, the electric part
of the 5D Weyl tensor with respect to the brane normal \cite{SMS}, averaged
over the two sides of the brane. The source term $\overline{L}_{ab}^{TF}$
originates in the asymmetric embedding of the brane and $\overline{\mathcal{P%
}}_{ab}$ is the pull-back of generic non-standard model fields in 5D \cite%
{Decomp}. The 4-dimensional (4D) and 5D gravitational coupling constants $%
\kappa ^{2}$ and $\widetilde{\kappa }^{2}$ are related as $6\kappa ^{2}=%
\widetilde{\kappa }^{4}\lambda $, with $\lambda $ the brane tension. The 4D
cosmological 'constant' $\Lambda $, apart from contributions of the
asymmetric embedding and non-standard model 5D fields, is defined as $%
2\Lambda _{0}=\kappa ^{2}\lambda +\widetilde{\kappa }^{2}\widetilde{\Lambda }
$.

Reference \cite{Decomp} gives the most generalic form of the gravitational
dynamics involving asymmetric embedding \cite{asymmetry} and non-static 5d
space-time due to radiation fields \cite{ChKN}-\cite{PalStructure}. Besides
cosmological applications in branes embedded into 5d black-hole space-times 
\cite{BCG}-\cite{BraneBlackHole} or into their horizon regions \cite%
{EinBrane}-\cite{EinBrane3} other aspects of brane-world models have been
discussed, including black-hole brane-worlds \cite{tidalRN}-\cite%
{BraneSwissCheese}, gravitational collapse on the brane \cite{BGM}-\cite%
{BraneOppSnyder}, stellar models \cite{GM}-\cite{Ovalle}, galactic dynamics 
\cite{HarkoRC}, the dynamics of clusters of galaxies \cite{HarkoClusters},
light deflection \cite{GeDa}, \cite{GeDefl} and Solar System tests \cite%
{BohmerHarkoLobo}.

A classical fluid membrane needs tension to exist. Similarly its higher
dimensional counterpart, the 3-brane, as it evolves, remains a hypersurface
due to the brane tension. The strongest bound on the minimal value of $%
\lambda $ was derived by combining the results of table-top experiments on
possible deviations from Newton's law, which probe gravity at sub-millimeter
scales \cite{tabletop} with the known value of the 4D Planck constant. In
the 2-brane model \cite{RS1} this gives \cite{Irradiated} $\lambda
>138.59\,\,$TeV$^{4}$. A much milder limit $\lambda \gtrsim 1$ MeV$^{4}$
arises from the constraint that the dominance of $S_{ab}$ ends before the
Big Bang Nucleosynthesis (BBN) \cite{nucleosynthesis}. From astrophysical
considerations on brane neutron stars an intermediate value $\lambda
>5\,\times 10^{8}$ MeV$^{4}$ was derived \cite{GM}. (All these limiting
values are for $c=1=\hbar $. In units $c=1=G$ the corresponding minimal
values of the brane tension are $\lambda _{tabletop}=4.\,\allowbreak 2\times
10^{-119}$ eV$^{-2}$, $\lambda _{BBN}=3\times 10^{-145}$ eV$^{-2}$ and $%
\lambda _{astro}=1.5\,\times 10^{-136}$ eV$^{-2}$, respectively \cite%
{BraneOppSnyder}.) For typical stellar densities the condition $\rho
_{star}/\lambda \ll 1$ is obeyed with any of these bounds.

In a cosmological context the brane represents our observable Universe.
Cosmic expansion is realized through the movement of the brane in the warped
extra dimension. During cosmological evolution the temperature of the brane
changes drastically. Cosmological branes cool down from a very hot early
Universe (whose thermal radiation is able to create a black hole in the
fifth dimension \cite{ChKN}), to the present days low temperature of the
Cosmic Microwave Background (CMB).

In Sec. II we explore the possibility of a variable brane tension, discussed
in detail in Ref. \cite{VarBraneTensionPRD}, by introducing a toy model, in
which the brane tension literally follows the temperature dependence of the
fluid tensions established for membranes. Such a model becomes particularly
simple when we assume the continuity equation. We discuss the emerging
cosmological model in Sec. III. A numerical solution of this toy-model,
which is compatible with cosmological observations is presented in Sec. IV.
Finally, Sec. V contains the concluding remarks.

Throughout the paper we follow the notations of Ref. \cite%
{VarBraneTensionPRD}.

\section{E\"{o}tv\"{o}s branes}

How justified is to assume a constant brane tension during cosmological
evolution, which spans over such a wide range of temperatures? The tension
of classical membranes depends on temperature, according to E\"{o}tv\"{o}s'
law \cite{Eotvos}%
\begin{equation}
\lambda _{fluid}=K\left( T_{c}-T\right) ~,  \label{EotvosLaw}
\end{equation}%
$K$ being a constant and $T_{c}\,$\ a critical temperature representing the
highest temperature for which the membrane exists.

Motivated by this analogy, the covariant gravitational dynamics on the brane
was analyzed in detail for variable tension brane-worlds, with the brane
asymetrically embedded into the 5d space-time, both the latter and the brane
containing arbitrary sources \cite{VarBraneTensionPRD}. After establishing
the covariant dynamics on the brane in a generic setup, a specialization to
a cosmological situation was presented, considering a Friedmann brane
asymmetrically embedded into a 5d space-time containing radiation.

As a first attempt to discuss in more detail the consequences of a
temperature-dependent brane tension, we adopt E\"{o}tv\"{o}s' law 
\begin{equation}
\lambda =\lambda _{lt}-\frac{6l}{\widetilde{\kappa }^{4}a}~,
\end{equation}%
where we have employed the standard relation \thinspace $T\propto a^{-1}$.
We have also denoted $KT_{c}=\lambda _{lt}$ and written the constant in the
second term in a suitable form, such that the 4D coupling 'constant' takes
the simple expression 
\begin{equation}
\kappa ^{2}=\kappa _{lt}^{2}-\frac{l}{a}~,
\end{equation}%
with 
\begin{equation}
\kappa _{lt}^{2}=\frac{\widetilde{\kappa }^{4}\lambda _{lt}}{6}.
\end{equation}%
The subscript $lt$ refers to late-time, as the second terms of both $\lambda 
$ and $\kappa ^{2}$ go to zero with$\ a\rightarrow \infty $.

Such a brane with temperature-dependent tension cannot exist below the
scale-factor $a_{\min }=l/\kappa _{lt}^{2}$ first because the tension would
become negative, leading to the destruction of the brane and secondly,
because the gravitational 'constant' would also become negative below this
limit, leading to anti-gravity on the brane. In terms of $a_{\min }$ the 4D
coupling 'constant' and the brane tension can be conveniently expressed as%
\begin{eqnarray}
\kappa ^{2} &=&\kappa _{lt}^{2}\left( 1-\frac{a_{\min }}{a}\right) ~,
\label{kappa2a} \\
\lambda  &=&\lambda _{lt}\left( 1-\frac{a_{\min }}{a}\right) ~.
\label{lambdaa}
\end{eqnarray}%
Both increase from zero to their asymptotic late-time values (Fig \ref%
{lambdakappa2a}). We also note that 
\begin{equation}
\widetilde{\kappa }^{4}=\frac{6\kappa ^{2}}{\lambda }=\frac{6\kappa _{lt}^{2}%
}{\lambda _{lt}}.
\end{equation}%
As the brane tension increases with scale factor according to the E\"{o}tv%
\"{o}s law, we may call this model an E\"{o}tv\"{o}s brane-world. The limits
derived for the brane-world tension from nucleosynthesis constraints in the
case of an E\"{o}tv\"{o}s brane refer to the value of the brane tension at
the time of nucleosynthesis, and in consequence its present-day value is
higher than for constant tension branes. 
\begin{figure}[tbp]
\includegraphics[height=4cm]{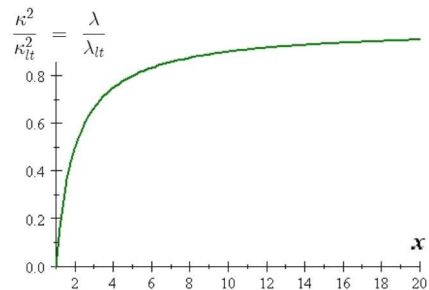}
\caption{(color online). The gravitational constant and brane tension
normalized to their late-time values ($\protect\kappa ^{2}/\protect\kappa %
_{lt}^{2}=\protect\lambda /\protect\lambda _{lt}$) represented as functions
of $x=a/a_{\min }$. Both the gravitational constant and brane tension are
zero at scale factor $a_{\min }$ and asymptotically increase to their
late-time values as $a\rightarrow \infty $.}
\label{lambdakappa2a}
\end{figure}

\begin{figure}[tbp]
\includegraphics[height=4cm]{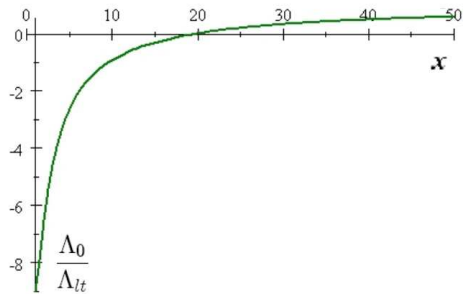}
\caption{(color online). The cosmological constant normalized to its
late-time value ($\Lambda _{0}/\Lambda _{lt}$), represented as function of $%
x=a/a_{\min }$ for the parameter value $L=1+\widetilde{\protect\kappa }^{2}%
\widetilde{\Lambda }/\protect\kappa _{lt}^{2}\protect\lambda _{lt}=2\Lambda
_{lt}/\protect\kappa _{lt}^{2}\protect\lambda _{lt}=0.1$ (The parameter $L$
obeys the inequalities $0<L<1$ due to the negativity of $\widetilde{\Lambda }
$ and the positivity of $\Lambda _{lt}$). The represented normalized
cosmological constant starts at high negative values, then it becomes
positive, increasing asymptotically to $1$ as $x\rightarrow \infty $.}
\label{Lambda0a}
\end{figure}

The $\Lambda _{0}$ contribution to the 4D cosmological 'constant' evolves cf.%
\begin{equation}
\Lambda _{0}=\Lambda _{lt}-\kappa _{lt}^{2}\lambda _{lt}\frac{a_{\min }}{a}%
\left( 1-\frac{a_{\min }}{2a}\right) ~,  \label{Lambda4a}
\end{equation}%
(Fig \ref{Lambda0a}) with its present day (late-time) value given by 
\begin{equation}
2\Lambda _{lt}=\kappa _{lt}^{2}\lambda _{lt}+\widetilde{\kappa }^{2}%
\widetilde{\Lambda }~.  \label{Lambda_lt}
\end{equation}%

As can be seen from Fig \ref{Lambda0a}, when the brane is formed at
temperature $T_{c}$, the contribution $\Lambda _{0}$ to the cosmological
'constant' is negative 
\begin{equation}
\Lambda _{c}=\Lambda _{lt}-\frac{\kappa _{lt}^{2}\lambda _{lt}}{2}=\frac{%
\widetilde{\kappa }^{2}\widetilde{\Lambda }}{2}<0.
\end{equation}%
Then, as the brane-world universe cools down with increasing$\ a$, the
factor in the second term of Eq. (\ref{Lambda4a}) obeys%
\begin{equation}
\frac{d}{da}\left[ \frac{a_{\min }}{a}\left( 1-\frac{a_{\min }}{2a}\right) %
\right] =-\frac{a_{\min }}{a^{2}}\left( 1-\frac{a_{\min }}{a}\right) <0~.
\end{equation}%
Therefore the second term of Eq. (\ref{Lambda4a}) is positive, resulting in
an increasing $\Lambda _{0}$ throughout the cosmological evolution, from $%
\Lambda _{c}<0$ to a positive $\Lambda _{lt}$ for $a\rightarrow \infty $,
obeying $\Lambda _{lt}\ll -\Lambda _{c}$.

These features imply the following modifications on the physics of the early
brane-world universe, first discussed in Ref. \cite{BDEL}: (a) brane-world
effects for an E\"{o}tv\"{o}s brane are more dominant then for a constant
tension brane, due to the initial smallness of the brane tension (this also
implies that the typical brane-world source term $\rho ^{2}/\lambda $,
arising from $S_{ab}$, dominates for a longer time); (b) due to the initial
smallness of $\kappa ^{2}$, gravity is initially quite weak; and (c) the
huge negative value of the cosmological constant generates an apparent
gravitational attraction.

In order to have a small $\Lambda _{lt}$, the values of $\lambda _{lt}$ and $%
\widetilde{\Lambda }$ have to be almost perfectly fine-tuned. As the
astrophysical lower limit refers to $\lambda _{lt}$, we can safely assume
the usual high negative value for the 5D cosmological constant $\widetilde{%
\kappa }^{2}\widetilde{\Lambda }$. In consequence the initial 4D
cosmological constant $\Lambda _{c}$ and its late-time value $\Lambda _{lt}$
obey $-\Lambda _{c}\gg \Lambda _{lt}$. Therefore the 4D cosmological
'constant' at early times represents a huge contribution in the balance of
sources.

\section{Cosmology}

According to the Stefan-Boltzmann law the energy density of the CMB (which
defines the temperature $T$) is proportional to the fourth power of $T$, and
further, according to the assumption $T\propto a^{-1}$, to $a^{-4}$. This is
possible only if the continuity equation holds.

Due to the variable brane tension and possible existence of a non-standard
model energy-momentum tensor $\widetilde{T}_{cd}$ in the fifth dimension,
the energy density of the cosmological fluid however obeys a more
sophisticated balance equation \cite{VarBraneTensionPRD}:%
\begin{equation}
\dot{\rho}+3\frac{\dot{a}}{a}\left( \rho +p\right) =-\dot{\lambda}+\Delta
\left( u^{c}{}n^{d}{}\widetilde{T}_{cd}\right) ~.  \label{f_en_balance_time}
\end{equation}%
Here $u$ is the 4-velocity of the fluid flow-lines, $\Delta $ denotes the
difference taken on the right and left sides of the brane, while a dot
represents the derivative with respect to cosmological time $\tau $. Note
that the normal vectors on the two sides of the brane are $n_{R}=n$ and $%
n_{L}=-n$, therefore the second term on the right hand side of Eq. (\ref%
{f_en_balance_time}) can be non-vanishing even in the symmetric case. In
order to have a continuity equation on the brane, thus the condition 
\begin{equation}
\dot{\lambda}=\Delta \left( u^{c}{}n^{d}{}\widetilde{T}_{cd}\right)
\label{lambdadot}
\end{equation}%
should hold. For an expanding (collapsing) universe $\dot{\lambda}=\left(
d\lambda /da\right) \dot{a}>0$ ($<0$), therefore $\Delta \left(
u^{c}{}n^{d}{}\widetilde{T}_{cd}\right) >0$ ($<0$) as well, corresponding to
a brane absorbing energy from (radiating energy into) the 5D space-time.

Any 5D radiation field (in the geometrical optics approximation) has
non-vanishing projection $u^{c}{}n^{d}{}\widetilde{T}_{cd}$ \cite{Decomp},
such that 
\begin{equation}
\Delta \left( u^{c}{}n^{d}{}\widetilde{T}_{cd}\right) =\frac{3}{\widetilde{%
\kappa }^{2}a^{3}}\sum\limits_{I=L,R}\epsilon _{I}\left( -1\right) ^{\eta
_{I}}\beta _{I}\dot{v}_{I}^{2}
\end{equation}%
at the location of the brane $r=a$. Here the function 
\begin{equation}
\beta \left( v\right) =\epsilon \frac{dm}{dv}
\end{equation}%
is a measure of the (linear) energy density of radiation, $v$ is a null
coordinate, and $m\left( v\right) $ the mass function of the 5D space-time.
The 5D space-time is Vaidya-Anti de Sitter (VAdS5), with line element$~$%
\begin{eqnarray}
d\widetilde{s}^{2} &=&-f\left( v,r\right) dv^{2}+2\epsilon dvdr  \notag \\
&&+r^{2}\left[ d\chi ^{2}+\chi ^{2}\left( d\theta ^{2}+\sin ^{2}\theta d\phi
^{2}\right) \right] ~,
\end{eqnarray}%
where for a spatially flat brane%
\begin{equation}
f\left( v,r\right) =-\frac{2m\left( v\right) }{r^{2}}-\frac{\widetilde{%
\kappa }^{2}\widetilde{\Lambda }}{6}r^{2}~.  \label{f}
\end{equation}%
The radiation is ingoing (towards $r=0$) for $\epsilon =1$ and outgoing for $%
\epsilon =-1$, while $\eta $ takes the value $1$ if the region contains$\ r=0
$, and $0$ otherwise. (The null coordinate $v$ is outgoing for $\epsilon =1$
and ingoing for $\epsilon =-1$.) Therefore, for an energy-absorbing brane
the following combinations are allowed: ($\eta =1,~\epsilon =-1$) and ($\eta
=0,~\epsilon =1$), thus $\left( -1\right) ^{\eta }\epsilon =1$, while for a
radiating brane either ($\eta =1,~\epsilon =1$) or ($\eta =0,~\epsilon =-1$)
hold, thus $\left( -1\right) ^{\eta }\epsilon =-1$. The derivatives $\dot{v}$
and $\dot{r}$ are related as \cite{VarBraneTensionPRD} 
\begin{equation}
f\dot{v}=\epsilon \dot{r}+\left( -1\right) ^{\eta +1}\left( \dot{r}%
^{2}+f\right) ^{1/2}~,  \label{vdotrdotf}
\end{equation}%
the sign of $\dot{v}$ being given by $\left( -1\right) ^{\eta +1}$. From
Eqs. (\ref{lambdaa}), (\ref{lambdadot}) and the relation between $\beta $
and $m$, by defining $\left( dm/dv\right) \dot{v}=\dot{m}$ (which allows to
introduce $m\left( \tau \right) $ in the equation), finally inserting the
expression (\ref{vdotrdotf}) evaluated on the brane in place of the
remaining factor $\dot{v}$, we find%
\begin{equation}
2a\dot{a}=\frac{\widetilde{\kappa }^{2}}{\kappa _{lt}^{2}a_{\min }}%
\!\sum\limits_{I=L,R}\!\!\dot{m}_{I}\frac{\epsilon _{I}\left( -1\right)
^{\eta _{I}}\dot{a}-\left( \dot{a}^{2}+f_{I}\right) ^{1/2}}{f_{I}}~.
\label{cond}
\end{equation}%
We chose the simplest case of a symmetrically embedded brane. By employing
the identity 
\begin{equation}
\epsilon \left( -1\right) ^{\eta }\dot{a}-\!\left( \dot{a}^{2}+f\right)
^{1/2}=-f\left[ \!\epsilon \left( -1\right) ^{\eta }\dot{a}+\left( \dot{a}%
^{2}+f\right) ^{1/2}\!\right] ^{-1},
\end{equation}%
Eq. (\ref{cond}) can be rewritten as 
\begin{equation}
\frac{\!\dot{m}}{a^{2}}\!=-\!\left( \!\frac{\kappa _{lt}^{2}\lambda _{lt}}{6}%
\!\!\right) ^{1/2}\!\frac{a_{\min }}{a}\dot{a}\left[ \epsilon \left(
-1\right) ^{\eta }\dot{a}+\left( \dot{a}^{2}+f\right) ^{1/2}\right] 
\label{cond1}
\end{equation}%
It is remarkable, that the above equation depends only on the combined sign $%
\epsilon \left( -1\right) ^{\eta }$. For a brane in the $f>0$ region Eq. (%
\ref{cond1}) implies $\dot{m}\dot{a}<0$.\ Indeed, the brane should absorb
(emit) radiation during expansion (contraction), and in consequence the mass
of the bounded 5D region decreases (increases). The positivity of the
radiation energy density $0<\beta \left( v\right) =\epsilon dm/dv=\epsilon 
\dot{m}\dot{v}^{-1}$ implies $\epsilon \left( -1\right) ^{\eta +1}sgn(\dot{m}%
)>0$, confirming $\epsilon \left( -1\right) ^{\eta }=1$ during expansion and 
$\epsilon \left( -1\right) ^{\eta }=-1$ during contraction.

The Friedmann equation \cite{Decomp}, 
\begin{equation}
\frac{\dot{a}^{2}}{a^{2}}=\frac{\Lambda _{0}}{3}+\frac{\kappa ^{2}\rho }{3}%
\left( 1+\frac{\rho }{2\lambda }\right) +\frac{2m}{a^{4}},
\end{equation}%
is not affected by the assumption of a variable brane tension \cite%
{VarBraneTensionPRD}, and can be used to eliminate $\dot{a}$ from the right
hand side of Eq. (\ref{cond1}). By inserting the $a$-dependent expressions
of $\lambda ,$~$\kappa ^{2}$ and $\Lambda _{0}$, the Friedmann equation
becomes%
\begin{eqnarray}
\frac{\dot{a}^{2}}{a^{2}} &=&\frac{\Lambda _{lt}}{3}\!+\frac{\kappa
_{lt}^{2}\rho }{3}\!\left( 1+\!\frac{\rho }{2\lambda _{lt}}\right) +\frac{2m%
}{a^{4}}  \notag \\
&&-\frac{\kappa _{lt}^{2}\lambda _{lt}}{3}\frac{a_{\min }}{a}\left( 1+\frac{%
\rho }{\lambda _{lt}}-\frac{a_{\min }}{2a}\right) ~.  \label{Fr_Va_varla}
\end{eqnarray}%
The last term represents first and second order corrections in $a_{\min }/a$
to the constant tension brane-world Friedmann equation. We have checked that
the Raychaudhuri equation and twice-contracted Bianchi identity are
consequences of Eqs. (\ref{f_en_balance_time}) and (\ref{Fr_Va_varla}). 
\begin{figure}[tbp]
\includegraphics[height=4.5cm]{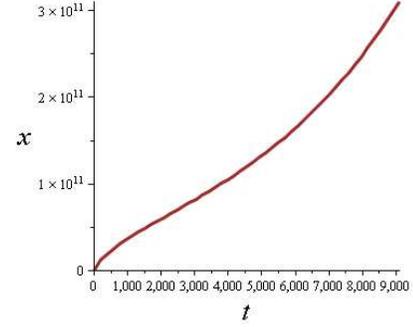}
\caption{(color online). The evolution of the scale factor for matter
dominated universe. The initial decelerated expansion is followed by an
ever-accelerating phase. (Plot for $R=10^{25},~y_{c}=10^{9}R$ and $%
L=4R/x_{0}^{3}$, with the inflection point at about $x_{0}\approx 10^{11}$.)}
\label{aevol}
\end{figure}
\begin{figure}[tbp]
\includegraphics[height=4.5cm]{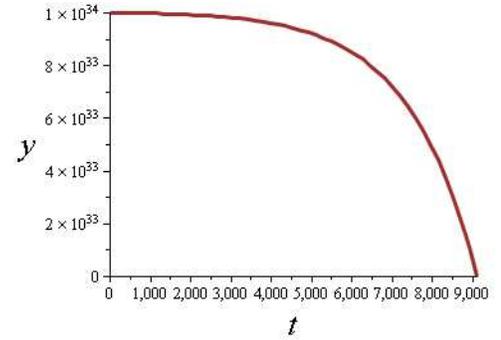}
\caption{(color online). The mass function of the 5D space-time decreases
until reaching maximal symmetry (AdS5). Parameters as for Fig. \protect\ref%
{aevol}.}
\label{mevol}
\end{figure}

\section{Numerical solution}

The continuity equation gives $\rho =\rho _{c}\left( a_{\min }/a\right) ^{n}$
with $n=3$ for matter and $n=4$ for radiation. Here $\rho _{c}$ is the
density at the creation of the brane. Then, by denoting $\mathcal{T}%
^{2}=6/\kappa _{lt}^{2}\lambda _{lt}$, we introduce the following
dimensionless variables 
\begin{eqnarray}
L &=&\frac{\Lambda _{lt}\mathcal{T}^{2}}{3}~,\quad R=\frac{\rho _{c}}{%
\lambda _{lt}}~,  \notag \\
x &=&\frac{a}{a_{\min }}~,\quad y=\frac{m\mathcal{T}^{2}}{a_{\min }^{4}}%
~,\quad t=\frac{\tau }{\mathcal{T}}~.
\end{eqnarray}%
The evolution [given by the system of Eqs. (\ref{Fr_Va_varla}) and (\ref%
{cond1})] of the dimensionless variables $x$ and $y$, in terms of the
dimensionless time parameter $t$ (the derivative with respect to $t$ being
denoted by a prime, and employing Eqs. (\ref{Lambda_lt}), (\ref{f}) in the
process), becomes:%
\begin{eqnarray}
x^{\prime }{}^{2} &=&1-2x+L\!x^{2}  \notag \\
&&+\frac{R}{x^{n-2}}\left( 2-\frac{2}{x}+\!\frac{R}{x^{n}}\right) +\frac{2y}{%
x^{2}}~,  \label{xdot} \\
\frac{y^{\prime }}{x} &=&\epsilon \left( -1\right) ^{\eta +1}x^{\prime
2}-x^{\prime }\left[ \!x^{\prime 2}\!\!+\!\!\left( 1-\!\!L\right)
x^{2}\!\!-\!\!\frac{2y}{x^{2}}\!\right] ^{1/2}.  \label{ydot}
\end{eqnarray}%
The variable $x$ increases from $1$ and its present day value is $%
x_{0}=z_{\max }+1\gg z_{BBN}\approx 4.26\times 10^{9}$ (where $z_{\max }$
corresponds to $a_{\min }$). The parameters of the model obey 
\begin{equation}
0<L\ll 1  \label{c1}
\end{equation}%
(from the positivity and smallness of $\Lambda _{lt}$, compared to any of
the $\kappa _{lt}^{2}\lambda _{lt}$, $-\widetilde{\kappa }^{2}\widetilde{%
\Lambda }$). From the dominance at present day of the $\rho $-term over the
correction terms containing $a_{\min }/a$ and over the $\rho ^{2}/2\lambda
_{lt}$ term of the Friedmann equation we obtain 
\begin{equation}
x_{0}^{2}\ll R\ll x_{0}^{3}~.  \label{c2}
\end{equation}%
Since today there is approximately twice as much dark energy (represented by 
$\Lambda _{lt}$) as matter, 
\begin{equation}
L\approx 4R/x_{0}^{3}~.  \label{c3}
\end{equation}%
The present day contribution of the mass term to the Hubble expansion being
also small \cite{supernova}, the condition 
\begin{equation}
y_{0}\ll Rx_{0}  \label{c4}
\end{equation}%
holds.

Numerical integration in this range of parameters gives an expanding
universe, with an initial decelerated phase followed by an accelerated
expansion (Fig \ref{aevol}). Due to the energy absorption of the brane the
mass of the VAdS5 region decreases (Fig \ref{mevol}). For the chosen
parameters at approximately three times the time when the dominance of $%
\Lambda _{lt}$ over matter begins, the mass $m\left( \tau \right) $ will
reach zero. With no mass left, the VAdS5 regions reduce to patches of 5D
Anti de Sitter (AdS5) space-time and the expansion on the brane continues in
a de Sitter phase.

\section{Concluding Remarks}

The variation of the brane tension introduces an additional degree of
freedom in brane-world models. The particular model discussed here, assuming
the E\"{o}tv\"{o}s law for the temperature dependence of the brane tension,
balanced by the energy interchange between the brane and VAdS5 (such that
the continuity equation holds), resulted in a monotonic increase with scale
factor of the brane tension, gravitational coupling constant and 4D
cosmological constant. In the early universe both the brane tension and the
4D gravitational coupling constant are small, enhancing the dominance of
brane-world effects. The temperature-dependent 4D cosmological constant,
being negative for small values of the scale factor, contributes to mutual
attraction, while positive for large $a$, generates dark energy type
repulsion.

We established the range of the model parameters allowed by the
confrontation with observations, given by Eqs. (\ref{c1})-(\ref{c4}). A
particular configuration obeying these conditions, with $%
R=10^{25},~x_{0}=10^{11},~y_{0}=10^{34}$ was represented on Figs \ref{aevol}%
, \ref{mevol}. For the allowed range the evolution of the fundamental
constants basically occur in the very early universe preceding BBN, after
which they asymptote to constant values. Still, feeded by absorbed energy
from the VAdS5 regions, they slightly evolve. This process eventually will
consume $m\left( \tau \right) $, leaving maximally symmetric AdS5 space-time
patches on the two sides of the brane. which further expands in a de Sitter
phase.

\section{Acknowledgements\textit{\ }}

This work has been successively supported by the Hungarian Scientific
Research Fund (OTKA) grant 69036, the J\'{a}nos Bolyai Grant of the
Hungarian Academy of Sciences, the London South Bank University Research
Opportunities Fund and the Pol\'{a}nyi Program of the Hungarian National
Office for Research and Technology (NKTH).

\end{document}